# A three-stage magnetic phase transition revealed in ultrahigh-quality van der Waals magnet CrSBr

Wenhao Liu[1*], Xiaoyu Guo[2*], Jonathan Schwartz[3], Hongchao Xie[2], Nikhil Dhale[1], Suk Hyun Sung[3], Aswin L. N. Kondusamy[1], Xiqu Wang[4], Haonan Zhao[2], Diana Berman[5], Robert Hovden[3,6#], Liuyan Zhao[2#], Bing Lv[1,7 #]

1. Department of Physics, the University of Texas at Dallas, Richardson, Texas 75080, USA
2. Department of Physics, University of Michigan, Ann Arbor, MI 48109, USA
3. Department of Material Science and Engineering, University of Michigan, Ann Arbor, MI 48109, USA
4. Department of Chemistry, University of Houston, Houston, Texas 77004, USA
5. Department of Materials Science and Engineering, University of North Texas, Denton, Texas 76203, USA
6. Applied Physics Program, University of Michigan, Ann Arbor MI 48109, USA
7. Department of Materials Science & Engineering, the University of Texas at Dallas, Richardson, Texas, 75080 USA.

## Abstract

van der Waals (vdW) magnets are receiving ever-growing attention nowadays due to their significance in both fundamental research on low-dimensional magnetism and potential applications in spintronic devices. High crystalline quality of vdW magnets is key for maintaining intrinsic magnetic and electronic properties, especially when exfoliated down to the 2D limit. Here, ultrahigh-quality air-stable vdW CrSBr crystals are synthesized using the direct vapor-solid synthesis method. The high single crystallinity and spatial homogeneity have been thoroughly evidenced at length scales from sub-mm to atomic resolution by X-ray diffraction, second harmonic generation, and scanning transmission electron microscopy. More importantly, specific heat measurements of these ultrahigh quality CrSBr crystals show three thermodynamic anomalies at 185K, 156K, and 132K, revealing a stage-by-stage development of the magnetic order upon cooling, which is also corroborated with the magnetization and transport results. Our ultrahigh-quality CrSBr can further be exfoliated down to monolayers and bilayers easily, paving the way to integrate them into heterostructures for spintronic and magneto-optoelectronic applications.

## Introduction

The advance of two-dimensional (2D) van der Waals (vdW) materials with thickness down to the atomic limit has opened a plethora of new research frontiers. The discoveries of 2D ferromagnetism with easy-axis anisotropy in vdW $Cr_2Ge_2Te_6$ and $CrI_3$ atomic layers have offered new platforms for studying 2D magnetism and new building blocks for integrating with spintronic and magneto-optoelectronic devices [1–3]. Sooner after, many other magnetic phases and tunable properties have been discovered in 2D magnets, both in their thin-film and bulk forms. For example, bulk $Fe_3GeTe_2$ (FGT) shows itinerant ferromagnetism with $T_c$ = 220-230 K[4,5], which can be enhanced to room temperature via ionic liquid gating in the monolayer[6]. $Cr_2Ge_2Te_6$ shows a semiconducting behavior and possesses a ferromagnetic transition temperature of 65 K in bulk [1,7], but can be tuned to 180 K in the few-layered samples when using solid ion conductors as the gate dielectric[8]. Chromium trihalides is another big family of 2D magnets. Whereas bulk $CrCl_3$ single crystals show in-plane antiferromagnetic (AFM) order below 17 K[9,10], CrBr$_3$ shows out-of-plane ferromagnetism below $T_c$ = 37 K [11] and CrI$_3$ hosts an out-of-plane ferromagnetic transition at about 61 K[12,13]. Besides, layer-dependent magnetism has been found in CrCl$_3$ and CrI$_3$, from ferromagnetic (FM) in a monolayer to AFM in a few layers, and back to FM in bulk [14,15]. The halide atoms play an important role in determining the magnetic texture. Interestingly, as the Cr-Cr distance increases with the halogen size from Cl to Br to I, the direct exchange interaction is supposed to be weakened. This means that the superexchange mechanism, which prefers to form ferromagnetic alignment, plays a more important role in the chromium trihalides series[16]. In addition, from Cl to Br to I, the spin-orbit coupling (SOC) strength increases, resulting in the evolution of spin anisotropy from the easy-plan to isotropic to nearly isotropic to the easy-axis type[17]. Furthermore, by tuning the chemical ratio of halide elements, the easy axis can be tuned in CrCl$_{3-x}$Br$_x$ and frustrated magnetic regions are demonstrated in the phase diagram of CrCl$_{3-x-y}$Br$_x$I$_y$[18,19].

The weak interlayer coupling in 2D materials/magnets allows the vertical stacking of the same or different atomic layers, where exotic behaviors that are not accessible in individual layers may be introduced. For instance, the magnetic proximity effect probed by the Zeeman spin Hall effect through non-local measurements is observed in graphene/CrBr$_3$ heterostructures[20]. Large Anomalous Hall effect is observed in the Cr$_2$Ge$_2$Te$_6$/(Bi,Sb)$_2$Te$_3$ heterostructure, showing that exchange coupling between the two different layers is strong and a sizable exchange gap opens in the surface state[21]. Antisymmetric magnetoresistance effect is observed in the Fe$_3$GeTe$_2$/Graphite/ Fe$_3$GeTe$_2$ trilayer heterostructures, which is proposed to result from a spin-polarized current induced by spin-momentum locking at the graphite/FGT interface[22]. Similar antisymmetric magnetoresistance is observed in the antiferromagnetic/ferromagnetic heterostructure of MnPS$_3$/Fe$_3$GeTe$_2$[23]. More recently, new magnetic states are discovered in moiré superlattices of twisted CrI$_3$ layers[24–26]. However, most of the 2D magnets have an out-of-plane easy axis, which are severely affected by the vertical heterostructure integration. Moreover, many of the 2D magnets are limited to their low magnetic transition temperature and the research of atomic thin *vdW* magnets is often hindered by their extreme sensitivity to air and degraded properties during the fabrication process. Thus, the pursuit of new 2D magnetic materials with preferably an in-plane magnetic easy axis, better crystal and exfoliation qualities, higher air stability, and better interfacial properties will be rather rewarding in the 2D magnets studies.

Very recently, an air-stable 2D antiferromagnetic CrSBr with a high Néel temperature $T_N$ =132 K has been reported[27,28]. As shown in Figure 1a, the Cr-S framework is sandwiched by Br atoms, forming one layer of CrSBr. Both Cr and S are octahedrally-coordinated and each layer of CrSBr is weakly coupled via the vdW force along *c* direction, resulting in the easy exfoliation of this material. Besides the high magnetic transition temperature, the single layer of CrSBr shows a ferromagnetic order with spin orientated along the in-plane *a*-axis, and these FM layers couple antiferromagnetically along the *c*-axis. The monolayer is also demonstrated to be air-stable and shows an in-plane ferromagnetic order below 146 K based on the second harmonic generation (SHG) results[29]. Furthermore, a strong spin-charge coupling is observed in the graphene/CrSBr heterostructure, which is due to a large induced exchange interaction by the proximity of the antiferromagnetic CrSBr[30]. To further facilitate the future heterostructure and device studies of this material, a high-quality bulk crystal which enables a better and large-area exfoliation will be necessary. Currently, the CrSBr are usually synthesized through the chemical vapor transport (CVT) method using chromium metal and S$_2$Br$_2$ sealed in quartz ampoule in two-zone tube furnaces [27,28]. However, vdW materials grown from CVT method often present more defects than other methods[31,32], which potentially impacts the magnetic order in 2D. Besides, here S$_2$B$_2$ plays a joint role of transport agent and reactants, which would cause the crystal easily cluster together and hinder the size and quality of the single crystal. Herein we present a simpler and more direct solid-vapor synthesis method. The high single crystallinity and homogeneity of the single crystals have been thoroughly evidenced at different length scales by X-ray diffraction (XRD), second harmonic generation rotational anisotropy (SHG RA), and scanning transmission electron microscopy (STEM), respectively.

## Results and Discussion

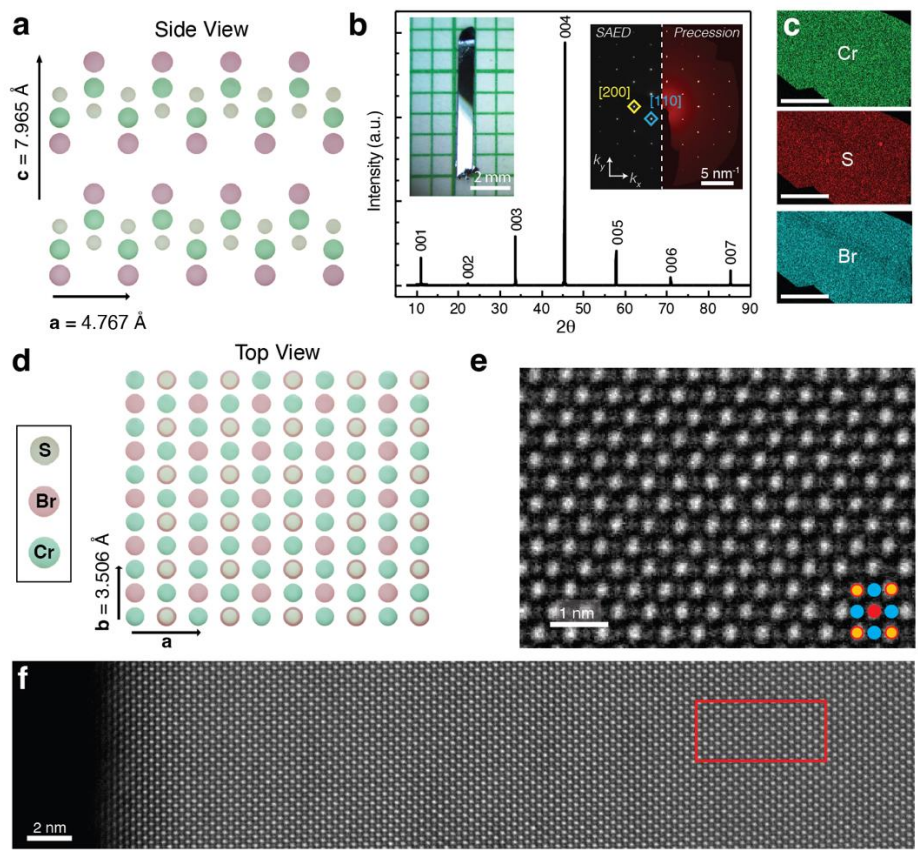

**Figure 1.** (a,d) Crystal structures of CrSBr viewed from *b* – and *c*– axis. (b) X-ray diffraction pattern with Miller indices on CrSBr single crystals. The left inset shows the optical image of the CrSBr single crystal on the millimeter grid. The right inset shows a selective area electron diffraction pattern along [001] and a precession image of (hk0) zone constructed by using a set of 1209 measured ω-scan frames. (c) EDX mapping of CrSBr, suggesting uniform distributions of constitute elements throughout the crystal. Scale bar, 250 μm. (e,f) HAADF-STEM images of the CrSBr crystal highlighting the presence of minimal defects over a 35 nm field-of-view.

Figure 1b shows the optical image and the XRD pattern of the CrSBr single crystal. The black and shiny single crystals with a lateral size as large as 7mm × 1 mm are typically obtained (More optical images in Figure S4). Only *00l* peaks can be observed as expected since the flat surface is perpendicular to the crystallographic *c*-axis. The full width at half maximum of the peaks is as narrow as 0.04º, which indicates the large crystalline domains and no noticeable inhomogeneous strain fields within samples. Procession X-ray and electron diffraction (Fig. 1b, inset) clearly demonstrate single-crystal nature with sharp Bragg peaks without distortions such as ring shape or blurred spots, suggesting near-perfect single crystallinity without noticeable grain boundaries or distortions in the samples over ~ 100 µm length scale. This is also reflected by the nice refinement results ($R_1$= 1.35% and $wR_2$= 3.51%, refinement details shown in Table S1). Furthermore, we construct procession images along the three (0kl), (h0l), (hk0) plane directions pixel-by-pixel from the complete sets of single-crystal XRD ω-scan images (1209 frames). As the precession images have detailed information of diffraction layers in the reciprocal space, it provides a direct and clear viewpoint on the crystal quality. The three procession images, shown in Figure 1c and Figure S1, further demonstrate no noticeable crystal twining or lattice distortion in three dimensions in our single-crystal samples. The lattice parameters of CrSBr at 170 K and 296 K are shown in Table S1.

We further demonstrate the high crystalline quality of CrSBr at the atomic scale in mechanical exfoliated atomic layers. Atomically resolved high-angle annular dark-field (HAADF) STEM measurements (Fig. 1e and 1f) confirm the absence of atomic defects, grain boundaries, or changes in stacking order in thin CrSBr flakes. The HAADF-STEM images show Cr and S-Br atomic columns viewed in projection as a cubic pattern with alternating dim (Cr) and bright spots (S-Br). The crystal uniformity is observed to the edge of the specimen (Fig. 1f). Furthermore, the selective area diffraction pattern (SAED) (Fig. 1b) reveals sharp Bragg spots with an in-plane two-fold symmetry, which agrees with the lattice periodicity of CrSBr.

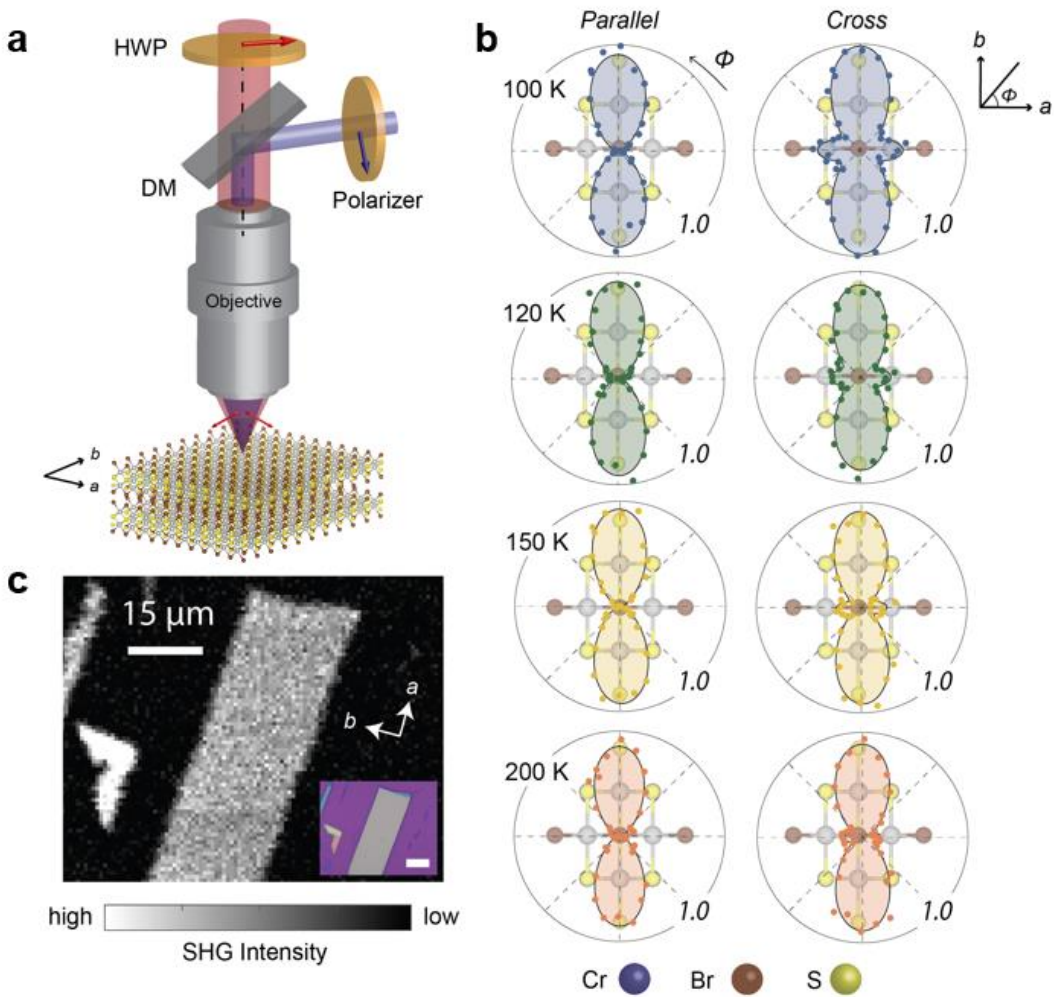

**Figure 2.** (a) Schematic of the SHG experimental setup. The incident fundamental and the reflected SHG light are in red and blue, respectively. Red arrows indicate the incident light polarizations while the blue arrow indicates the selected reflected SHG light polarization. HWP: half-wave plate, DM: dichroic mirror. (b) Polar plots of SHG RA data in both the parallel and the crossed channels at selected temperatures. The experiment data (filled circles) are well fitted by the model based on group theory analysis (solid lines). All plots are of the same intensity scale, normalized to a value of 1.0 corresponding to 35 pW. The CrSBr crystal structures are overlaid in the background. Inset shows the definition of polar angle $\phi$ that is the angle between the incident light polarization and the crystal axis $a$. (c) Scanning SHG image of the bulk CrSBr sample measured at $\phi = 90°$ in the parallel channel at 80 K. Inset shows the corresponding wide-field optical image. The scale bar of the inset is 15 μm.

Secondly, to further characterize the homogeneity of the single crystal in the micron length scale, we have carried out SHG RA measurements on bulk CrSBr under the nominal normal incidence geometry. Figure 2a shows the schematic of the experimental setup. A long working distance $20 \times$ objective with a 0.4 numerical aperture (NA) focuses the 800 nm incident laser beam onto the sample surface. The relatively large NA yields a cone-shaped focused beam, covering the incident electric field polarizations from perfectly parallel to the sample surface to 23.5°, i.e., $\arcsin(NA)$, off from it. A dichroic mirror (DM) set, together with a bandpass filter set, selects the 400 nm SHG light and directs it to a photomultiplier-tube detector that is connected to a current amplifier and then a lock-in amplifier. The polarizations of the incident and the reflected light can be selectively chosen using a half-wave plate (HWP) and a polarizer, respectively.

We have performed SHG RA measurements to verify the structural symmetries at various temperatures across the AFM phase transition temperature $T_N = 132$ K. Figure 2b shows the polar plots of SHG intensity as a function of the incident polarization angle $\phi$ away from $a$-axis (Figure 2b inset) in two channels where the incident fundamental and the reflected SH light polarizations are parallel (*Parallel*) and perpendicular (*Cross*) to each other. We clearly see that all RA patterns show two mirrors normal to crystal axes $a$ and $b$, consistent with the point group *mmm* determined by our XRD measurement (space group *Pmmn*, No.59, see Table S1). Furthermore, we observe that from 100 K to 200 K, the SHG RA results in the parallel channel remains nearly constant including both the shape and the amplitude whereas those in the crossed channel shows a modest evolution. The two smaller lobes in SHG RA patterns of the crossed channel along the $a$-axis gradually shrink as the temperature increases, while the two larger lobes along the $b$-axis barely change in intensity. The absence of significant changes in both channels across $T_N = 132$ K is consistent with the same *mmm* point group for both the paramagnetic phase above $T_N$ and the layered antiferromagnetic phase below $T_N$, demonstrating the robust and consistent crystal structure of the samples. Usually, the 2-fold rotation symmetry operation along the $c$-axis should forbid any SHG radiation in the normal incidence geometry, including both the electric dipole and the electric quadrupole contributions. Our observed non-zero SHG RA signal is attributed to the electric quadrupolar contribution from the large NA-induced oblique light rays (Figure 2a). We further simulated the SHG RA functional form by computing the electric quadrupolar SHG under the *mmm* point group and then taking an average over the full azimuthal direction at an oblique angle $\theta = \arcsin(NA)$ [See Supplementary information]. The experimental data is well fitted by this model as shown in Figure 2b.

In addition to the single-spot SHG RA measurement to confirm the point symmetry, we have also performed scanning SHG microscopy measurements to examine the spatial homogeneity at a spatial resolution of sub-μm and a length scale of several tens of μm. Figure 2c shows an SHG scanning image measured at $\phi = 90°$ in the parallel channel at 80 K, showing a uniform distribution of the SHG intensity, further confirming the structural and magnetic homogeneity of the sample at micron scale.

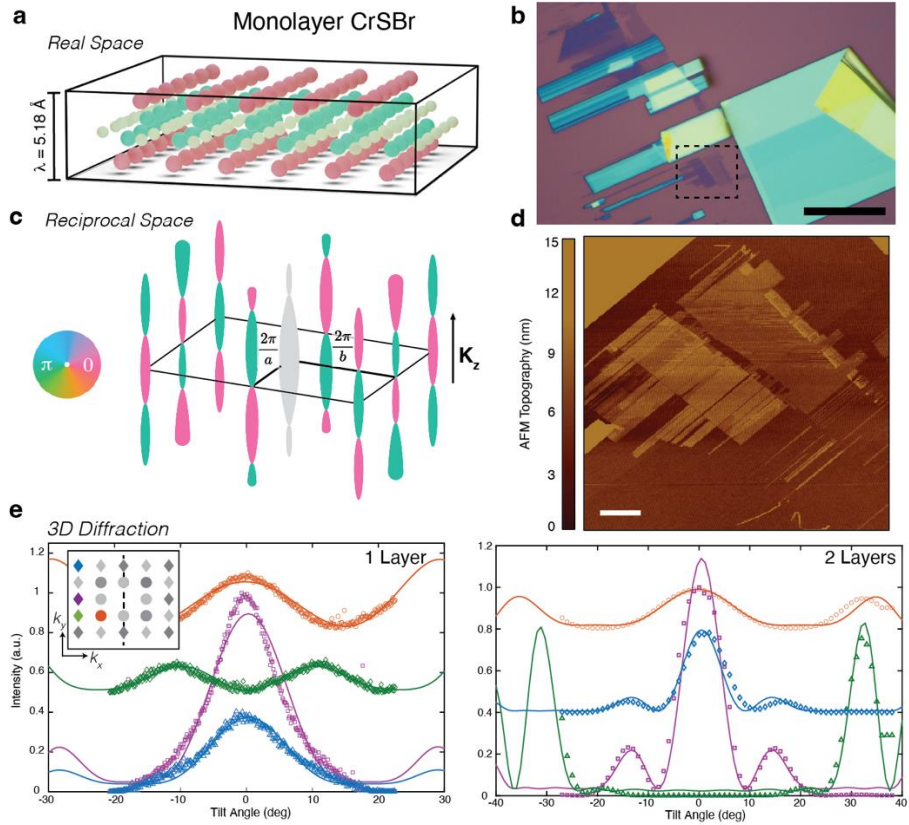

**Figure 3**. (a) Schematic illustration of monolayer CrSBr. (b) Optical image of thin CrSBr flakes deposited onto a SiO$_2$/Si wafer substrate via mechanical exfoliation. (c) Schematic of calculated out-of-plane momentum ($k_z$) dependence for various Bragg rods of monolayer CrSBr. The thickness and color indicate the complex magnitude and phase of the structure factor, respectively. (d) Atomic force microscopy topography of exfoliated CrSBr flakes. (e) Experimental Bragg intensities (scatter points) for ($\bar{1}\bar{1}0$), ($\bar{2}20$), ($\bar{2}00$), and ($\bar{2}\bar{1}0$) peaks plotted as a function of tilt angle, which strongly agree with the fitted kinematic models.

Monolayer (Fig. 3a) and bilayer are readily fabricated as large, single crystals from the bulk using mechanical exfoliation, as confirmed with AFM and 3D reciprocal space measurements[33]. Figure 3d shows an AFM image of monolayer and bi-layer CrSBr flakes, which can be routinely obtained by mechanical exfoliation from bulk single crystals. By tilting the sample, we can measure the diffraction spots as a function of out-of-plane crystal momentum ($k_z$), A schematic of the 3D Bragg structure for several monolayer CrSBr peaks are shown in Figure 1c. The experimental Bragg rod intensities are shown in Figure 1e as discrete points together with their expected values in solid lines. As the $k_z$ dependence for bilayer flakes are noticeably different, we can confirm our ability to exfoliate CrSBr crystals down to monolayer thickness. In addition, fitting the $k_z$ dependence for the Bragg peaks provides direct quantification for the out-of-plane spacing between two adjacent elements. Specifically we experimentally measured ΔCr = 0.90 ±0.16 Å, ΔS = 0.61±0.09 Å, ΔBr = 2.59 ± 0.19 Å and c = 7.91 ± 0.288 Å for the bilayer CrSBr crystal.

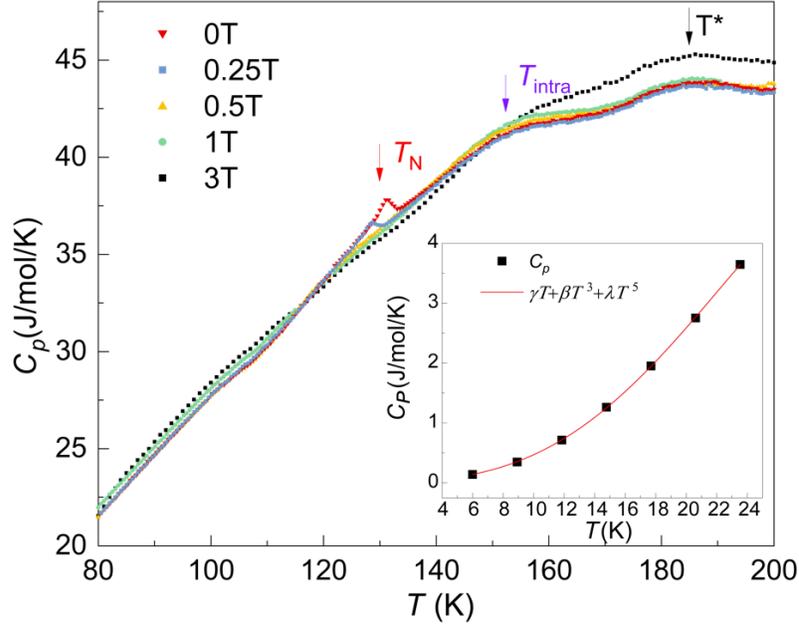

**Figure 4**. (a) Specific heat $C_p(T)$ at temperatures from 80 K to 200 K under various external magnetic fields applied along the $c$ direction of the CrSBr single crystal. Three specific anomalies are marked as $T_N$, $T_{intra}$ and $T^*$, respectively. The inset shows the fitting of the specific heat from 6 K to 24 K using the Debye model $C_p = \gamma T + \beta T^3 + \eta T^5$.

Given the exceptional crystalline quality of these CrSBr crystals, we revisit the thermodynamic properties of bulk CrSBr. Heat capacity measurement was conducted with applied magnetic fields parallel to the $c$-axis of the CrSBr single crystals. Low-temperature specific heat behaviors of CrSBr are shown in the inset of Figure 4, which can be fitted using the Debye model $C_p = \gamma_n T + \beta T^3 + \eta T^5$ where $\gamma_n T$ and $\beta T^3 + \eta T^5$ are from the electron and phonon contributions, respectively. The parameters obtained from the fitting results are $\gamma_n$ = 9.8 mJ/mol/K$^2$, $\beta$ = 0.396 mJ/mol/K$^4$, $\eta$ = -2.4×10$^{-4}$ mJ/mol/K$^6$. The Debye temperature $\Theta_D$ thus can be estimated using the relation $\Theta_D = \left(\frac{12\pi^4 k_B N_A Z}{5\beta}\right)^{1/3}$ to be $\Theta_D$ = 245 K, where $N_A$ is the Avogadro constant and $Z$ is the number of atoms in one unit cell. The nonzero $\gamma_n$ in the semiconducting phase here may be associated with a constant density of states of antiferromagnetic excitations near the Fermi surface, which are discovered in many 2D semiconducting magnets[34].

In the high-temperature region, three distinct specific heat anomalies can be observed. From low to high temperatures, the first clear sharp peak is at 132 K at 0 T. Upon increasing magnetic field, this peak gets suppressed towards lower temperatures till disappearing above 0.5 T, which is a typical behavior of AFM phase transition happening at the Néel temperature $T_N$. The second hump occurs at ~155 K with a clear change of the slope. The transition temperature gets enhanced by applying magnetic fields, which is reminiscent of the intralayer ferromagnetic coherence (or order). The third anomaly, which has not been observed previously and is denoted as $T^*$, occurs around 185 K and remains nearly unchanged under external magnetic fields up to 3 T. It comes to our attention that, in the isostructural compound CrOCl, similar three specific heat anomalies have also been observed in the high-quality crystals. CrOCl shows an antiferromagnetic order below $T_N = 13.5$K. At 26.7, 27.8K, two other anomalies emerge, which are related to the formation of the incommensurate magnetic superstructure and a structure structural phase transition[35,36]. However, for our crystals, low-temperature single-crystal X-ray diffraction at 170 K (Table S1) and SHG RA measurements (Figure 2) do not show any symmetry change nor structural transition in the bulk crystals.

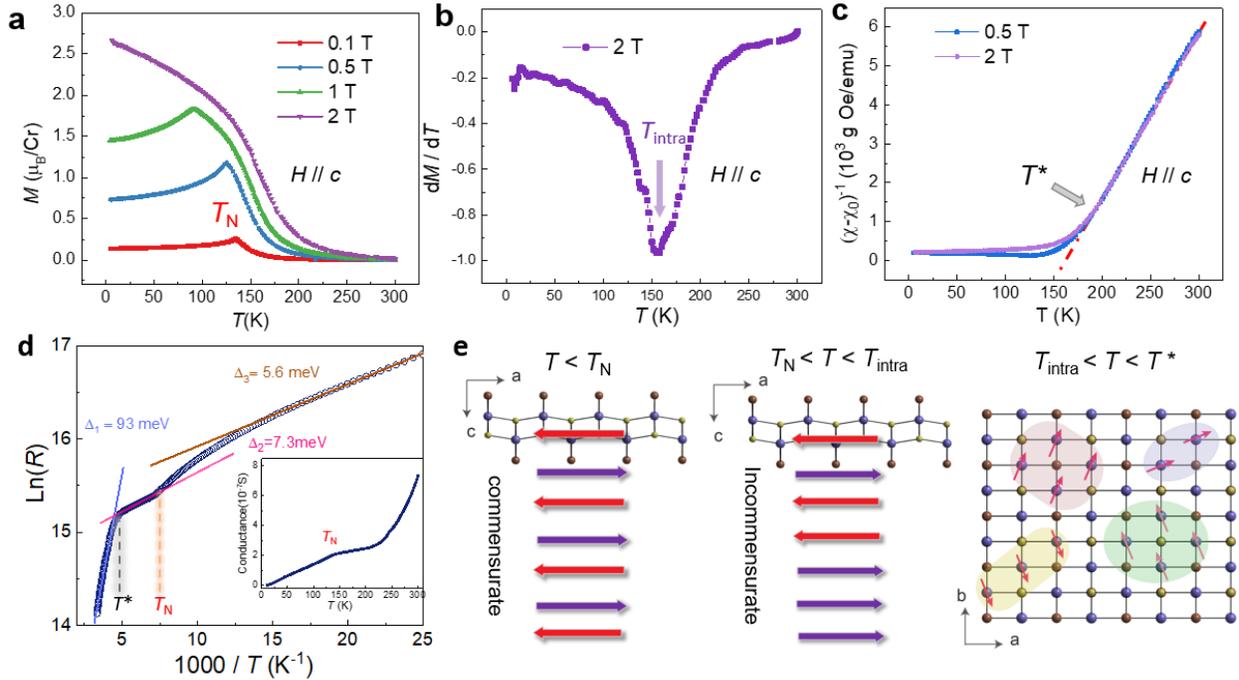

**Figure 5 Magnetization and transport behaviors of CrSBr**. (a) Temperature-dependent magnetization with various magnetic fields applied parallel to $c$-axis. (b) dM/dT curve of the magnetization under the external magnetic field of 2T, where the antiferromagnetic order is completely suppressed, indicating a phase transition at 155 K. (c) Temperature dependence of the inverse magnetic susceptibility with the magnetic field of 0.5 T and 2 T applied parallel to the $c$-axis. The red line shows the fitting of our data using the Curie-Weiss law, where a deviation occurs around 185 K. (d) Ln(R) vs. 1/T. Transport behaviors are fitted using the thermal activation model. The gap changes across the specific heat anomalies occurring temperatures. Inset: temperature dependence of in-plane conductivity of CrSBr. (e) Schematic magnetic structures of CrSBr upon cooling. The ellipses here mean the local magnetic order developed.

To investigate further these thermodynamic anomalies, magnetization together with transport measurements have been carried out. Figure 5a exhibits the temperature dependencies of magnetizations for bulk CrSBr single crystals in the magnetic fields ranging from 0.1 to 2 T applied along $c$-axis. A peak around 132 K under 0.1 T stands out clearly. Upon increasing the external magnetic field, this peak becomes broader and shifts to lower temperatures and at last disappears above 2 T, which is a classic behavior for an antiferromagnetic phase transition consistent with the specific heat anomaly at $T_N$. Besides, the temperature dependence of magnetizations with fields applied along $a$-axis is shown in Figure S2, where the AFM order is totally suppressed above 1 T. The magnitudes of the magnetic field that suppress the AFM order along $a$- and $c$-axis are very different, demonstrating strong magnetic anisotropy in CrSBr. In addition, in the low-temperature region of Figure 5a, the magnetization measured along the *c*-axis saturates with a value of 2.75 $\mu_B$ /Cr, which is very close to the expected value of 3 $\mu_B$ expected for the trivalent Cr with an electronic configuration $3d^3$, indicating nearly all the moments are ferromagnetically aligned along the *c*-axis. It is worthwhile noting that a recent report of a magnetization hump around 40K observed in their CrSBr sample, which is attributed to a hidden low-temperature magnetic order.[37] However, such magnetization hump is absent from the magnetization measurements on our high-quality samples.

Figure 5b plots the temperature dependence of dM/dT under the magnetic field of 2 T along the *c*-axis when the long-range AFM order is completely suppressed. A clear dip, indicating a phase transition near 155 K can be clearly seen and is consistent with the specific heat anomaly occurred at $T_{intra}$. This phase transition is attributed as the occurrence of the intralayer superexchange coupling and an intermediate ferromagnetic phase (iFM)[29] where the ferromagnetic order within the plane is established while the interlayers show an incommensurate magnetic order.

The inverse magnetic susceptibilities under 0.5 T and 2 T are shown in Figure 5c and demonstrate the deviations from the Curie-Weiss behavior at ~185 K. The red line in Figure 5c is the fitting result by using inverse magnetic susceptibilities through the modified Curie-Weiss law

$$\chi = \chi_0 + \frac{C}{T - T_0}$$

where $\chi_0$ is the temperature-independent susceptibility arising from the background, *C* is the Curie-Weiss constant and $T_0$ is the Curie temperature. The yielding values are $\chi_0$= -1.4 $\times 10^{-4}$ emu/g/Oe, C= 0.0226 emu K /g /Oe, $T_0$=164 K for the magnetization under 0.5 T and $\chi_0$= -1.4 $\times 10^{-4}$ emu/g/Oe, C= 0.0231 emu K /g /Oe, $T_0$=163 K or the magnetization under 2 T.

The temperature of the deviation from the Curie-Weiss behavior is close to *T\**, the third anomaly in the specific heat results. Previously we have demonstrated no structure or crystal symmetry changes across *T\**. Thus, this broad hump at *T\** in the specific heat could be related to the emergence of the spin-spin correlations. Upon increasing the magnetic field to 3T, the broad hump keeps nearly the same both in temperature and magnitude, which is reminiscent of that found in frustrated magnetic systems[38,39]. A plausible scenario here is that the short-range magnetic order begins to establish within each CrSBr layer while the long magnetic order within the layer is still prohibited due to strong thermal fluctuations. Only when the temperature decreases to $T_{intra}$ can the long magnetic order in each layer develop.

Those specific heat anomalies are also reflected in the transport results. The inset of Figure 5d shows the temperature dependence of the conductance of CrSBr. A clear change of the conductance slope can be observed around 132 K, which is attributed to the AFM phase transition. Figure 5d shows Ln(R) vs. (1000/T). Three linear relations, which are consistent with the standard thermal activation model[40], can be observed in different temperature regimes. The gap above *T\** is estimated to be about 92 meV and decreases dramatically to 7.3 meV for $T_N < T < T^*$. Such a drastic change of the gap reflects the strong coupling between the magnetic order and the transport behaviors in this material. The gap gets modified further for $T < T_N$ and finally decreased to 5.6 meV in the low-temperature region till 10 K. It is noted that $T_{intra}$ is not reflected in the transport results, which may be because the incommensurate magnetic order is established among adjacent layers at $T_{intra}$ while the temperature dependence of resistance here is measured within the layer of the CrSBr single crystal. Instead, the out-of-plane resistance measurement may reveal distinct features at $T_{intra}$, which needs future investigation.

Those specific heat anomalies together with the magnetization and transport results indicate that the magnetic order gradually develops in three different stages upon cooling, as is illustrated in Figure 5e. (1)

From room temperature to $T^*$, the crystal shows paramagnetic orders with random spin orientations. (2) For $T_{intra} < T < T^*$, local or short-range magnetic order begins to establish due to the spin-spin coupling within each CrSBr layer while the long magnetic order within the layer is still suppressed due to thermal fluctuations. When the temperature decreases to $T_{intra}$ where the thermal fluctuations are reduced, the long magnetic order in each layer is established. (3) For $T_N < T < T_{intra}$, an intermediate ferromagnetic phase (iFM) began to establish where the ferromagnetic order within the plane is established while the interlayers show incommensurate magnetic order. (4) For $T < T_N$, a ferromagnetic order within the plane together with antiferromagnetic order among adjacent planes is formed.

In conclusion, ultrahigh-quality air-stable and large-size bulk CrSBr single crystals have been synthesized directly using the vapor-solid growth method. The high single crystallinity and homogeneity of the single crystals have been thoroughly evidenced at different length scales by XRD, SHG RA, and STEM, respectively. More importantly, specific heat measurements of these ultrahigh quality CrSBr crystals show three thermodynamic anomalies at 185K, 156K, and 132K, revealing a stage-by-stage development of the magnetic order upon cooling, which is also corroborated with the magnetization and transport results. Our ultrahigh-quality CrSBr can further be exfoliated down to monolayers and bilayers easily, paving the way to integrate them into heterostructures for spintronic and magneto-optoelectronic applications.

## Acknowledgments

We thank Prof. Stephen Forrest for offering us to use the atomic force microscopy in his laboratory to calibrate layer numbers of exfoliated CrSBr. Work conducted at University of Texas at Dallas was supported by the U.S. Air Force Office of Scientific Research Grant No. FA9550-19-1-0037. This project is also partially supported by NSF- DMREF- 1921581 (magnetic characterizations). L.Z. acknowledges support by NSF CAREER grant no. DMR-174774, AFOSR YIP grant no. FA9550-21-1-0065, and Alfred P. Sloan Foundation. R.H. acknowledges support from ARO grant no. W911NF-22-1-0056 and the W.M. Keck Foundation. Support from the Advanced Materials and Manufacturing Processes Institute (AMMPI) at the University of North Texas is acknowledged.

## References

(1) Gong, C.; Li, L.; Li, Z.; Ji, H.; Stern, A.; Xia, Y.; Cao, T.; Bao, W.; Wang, C.; Wang, Y.; Qiu, Z. Q.; Cava, R. J.; Louie, S. G.; Xia, J.; Zhang, X. Discovery of Intrinsic Ferromagnetism in Two-Dimensional van Der Waals Crystals. *Nature* **2017**, *546*, 265–269.

(2) Huang, B.; Clark, G.; Navarro-Moratalla, E.; Klein, D. R.; Cheng, R.; Seyler, K. L.; Zhong, Di.; Schmidgall, E.; McGuire, M. A.; Cobden, D. H.; Yao, W.; Xiao, D.; Jarillo-Herrero, P.; Xu, X. Layer-Dependent Ferromagnetism in a van Der Waals Crystal down to the Monolayer Limit. *Nature* **2017**, *546*, 270–273.

(3) Seyler, K. L.; Zhong, D.; Klein, D. R.; Gao, S.; Zhang, X.; Huang, B.; Navarro-Moratalla, E.; Yang, L.; Cobden, D. H.; McGuire, M. A.; Yao, W.; Xiao, D.; Jarillo-Herrero, P.; Xu, X. Ligand-Field Helical Luminescence in a 2D Ferromagnetic Insulator. *Nat. Phys.* **2018**, *14*, 277–281.

(4) Chen, B.; Yang, J. H.; Wang, H. D.; Imai, M.; Ohta, H.; Michioka, C.; Yoshimura, K.; Fang, M. H. Magnetic Properties of Layered Itinerant Electron Ferromagnet $Fe_3GeTe_2$. *J. Phys. Soc. Japan* **2013**, *82* 124711.

(5) May, A. F.; Calder, S.; Cantoni, C.; Cao, H.; McGuire, M. A. Magnetic Structure and Phase Stability of the van Der Waals Bonded Ferromagnet $Fe_{3-x}GeTe_2$. *Phys. Rev. B* **2016**, *93*, 014411.

(6) Deng, Y.; Yu, Y.; Song, Y.; Zhang, J.; Wang, N. Z.; Wu, Y. Z.; Zhu, J.; Wang, J.; Chen, X. H.; Zhang, Y. Gate-Tunable Room-Temperature Ferromagnetism in Two-Dimensional $Fe_3GeTe_2$. *Nature* **2018**, *563*, 94-99.

(7) Carteaux, V.; Brunet, D.; Ouvrard, G.; Andre, G. Crystallographic, Magnetic and Electronic Structures of a New Layered Ferromagnetic Compound $Cr_2Ge_2Te_6$. *J. Phys. Condens. Matter* **1995**, *7*, 69–87.

(8) Zhuo, W.; Lei, B.; Wu, S.; Yu, F.; Zhu, C.; Cui, J.; Sun, Z.; Ma, D.; Shi, M.; Wang, H.; Wang, W.; Wu, T.; Ying, J.; Wu, S.; Wang, Z.; Chen, X. Manipulating Ferromagnetism in Few-Layered $Cr_2Ge_2Te_6$. *Adv. Mater.* **2021**, *33*, 2008586

(9) Vleck, V. Heat Capacities of $CrF_3$ and $CrCl_3$ from 15 to 300 °K. *J. Chem. Phys.* **1958**, *28*, 902–907.

(10) Elliott, N.; Hastings, J. Neutron Diffraction Investigation of KCN. *Acta Crystallogr.* **1961**, *14*, 1018–1018.

(11) Tsubokawa, I. On the Magnetic Properties of a $CrBr_3$ Single Crystal. *J. Phys. Soc. Japan* **1960**, *15*, 1664–1668.

(12) Dillon, J. F.; Olson, C. E. Magnetization, Resonance, and Optical Properties of the Ferromagnet $CrI_3$. *J. Appl. Phys.* **1965**, *36*, 1259–1260.

(13) McGuire, M. A.; Dixit, H.; Cooper, V. R.; Sales, B. C. Coupling of Crystal Structure and Magnetism in the Layered, Ferromagnetic Insulator $CrI_3$. *Chem. Mater.* **2015**, *27*, 612–620.

(14) Cai, X.; Song, T.; Wilson, N. P.; Clark, G.; He, M.; Zhang, X.; Taniguchi, T.; Watanabe, K.; Yao, W.; Xiao, D.; McGuire, M. A.; Cobden, D. H.; Xu, X. Atomically Thin CrCl3: An In-Plane Layered Antiferromagnetic Insulator. *Nano Lett.* **2019**, *19*, 3993–3998.

(15) Huang, B.; Clark, G.; Klein, D. R.; MacNeill, D.; Navarro-Moratalla, E.; Seyler, K. L.; Wilson, N.; McGuire, M. A.; Cobden, D. H.; Xiao, D.; Yao, W.; Jarillo-Herrero, P.; Xu, X. Electrical Control of 2D Magnetism in Bilayer $CrI_3$. *Nat. Nanotechnol.* **2018**, *13*, 544–548.

(16) Lado, J. L.; Fernández-Rossier, J. On the Origin of Magnetic Anisotropy in Two dimensional $CrI_3$. *2D Mater.* **2017**, *4*, 035002.

(17) Kim, H. H.; Yang, B.; Li, S.; Jiang, S.; Jin, C.; Tao, Z.; Nichols, G.; Sfigakis, F.; Zhong, S.; Li, C.; Tian, S.; Cory, D. G.; Miao, G. X.; Shan, J.; Mak, K. F.; Lei, H.; Sun, K.; Zhao, L.; Tsen, A. W. Evolution of Interlayer and Intralayer Magnetism in Three Atomically Thin Chromium Trihalides. *Proc. Natl. Acad. Sci. U. S. A.* **2019**, *166*, 11131.

(18) Tartaglia, T. A.; Tang, J. N.; Lado, J. L.; Bahrami, F.; Abramchuk, M.; McCandless, G. T.; Doyle, M. C.; Burch, K. S.; Ran, Y.; Chan, J. Y.; Tafti, F. Accessing New Magnetic Regimes by Tuning the Ligand Spin-Orbit Coupling in van Der Waals Magnets. *Sci. Adv.* **2020**, *6*, eabb9379

(19) Abramchuk, M.; Jaszewski, S.; Metz, K. R.; Osterhoudt, G. B.; Wang, Y.; Burch, K. S.; Tafti, F. Controlling Magnetic and Optical Properties of the van Der Waals Crystal $CrCl_{3-x}Br_x$ via Mixed Halide Chemistry. *Adv. Mater.* **2018**, *30*, 1801325

(20) Tang, C.; Zhang, Z.; Lai, S.; Tan, Q.; Gao, W. bo. Magnetic Proximity Effect in Graphene/CrBr3 van Der Waals Heterostructures. *Adv. Mater.* **2020**, *32,* 1908498.

(21) Mogi, M.; Nakajima, T.; Ukleev, V.; Tsukazaki, A.; Yoshimi, R.; Kawamura, M.; Takahashi, K. S.; Hanashima, T.; Kakurai, K.; Arima, T. H.; Kawasaki, M.; Tokura, Y. Large Anomalous Hall Effect in Topological Insulators with Proximitized Ferromagnetic Insulators. *Phys. Rev. Lett.* **2019**, *123,* 016804.

(22) Albarakati, S.; Tan, C.; Chen, Z.; Partridge, J. G.; Zheng, G.; Farrar, L.; Mayes, E. L. H.; Field, M. R.; Lee, C.; Wang, Y.; Xiong, Y.; Tian, M.; Xiang, F.; Hamilton, A. R.; Tretiakov, O. A.; Culcer, D.; Zhao, Y.; Wang, L. Antisymmetric Magnetoresistance in van Der Waals $Fe_3GeTe_2$/Graphite / $Fe_3GeTe_2$ Trilayer Heterostructures. *Sci. Adv.* **2019** *5*, eaaw0409.

(23) Hu, G.; Zhu, Y.; Xiang, J.; Yang, T. Y.; Huang, M.; Wang, Z.; Wang, Z.; Liu, P.; Zhang, Y.; Feng, C.; Hou, D.; Zhu, W.; Gu, M.; Hsu, C. H.; Chuang, F. C.; Lu, Y.; Xiang, B.; Chueh, Y. L. Antisymmetric Magnetoresistance in a van Der Waals Antiferromagnetic/Ferromagnetic Layered $MnPS_3$/$Fe_3GeTe_2$ Stacking Heterostructure. *ACS Nano* **2020**, *14*, 12037–12044.

(24) Xie, H.; Luo, X.; Ye, G.; Ye, Z.; Ge, H.; Sung, S. H.; Rennich, E.; Yan, S.; Fu, Y.; Tian, S.; Lei, H.; Hovden, R.; Sun, K.; He, R.; Zhao, L. Twist Engineering of the Two-Dimensional Magnetism in Double Bilayer Chromium Triiodide Homostructures. *Nat. Phys.* **2022**, *18*, 30–36.

(25) Xu, Y.; Ray, A.; Shao, Y. T.; Jiang, S.; Lee, K.; Weber, D.; Goldberger, J. E.; Watanabe, K.; Taniguchi, T.; Muller, D. A.; Mak, K. F.; Shan, J. Coexisting Ferromagnetic–Antiferromagnetic State in Twisted Bilayer $CrI_3$. *Nat. Nanotechnol.* **2021**.*17*, 143–147.

(26) Song, T.; Sun, Q. C.; Anderson, E.; Wang, C.; Qian, J.; Taniguchi, T.; Watanabe, K.; McGuire, M. A.; Stöhr, R.; Xiao, D.; Cao, T.; Wrachtrup, J.; Xu, X. Direct Visualization of Magnetic Domains and Moiré Magnetism in Twisted 2D Magnets. *Science.* **2021**, *374*, 1140–1144.

(27) Telford, E. J.; Dismukes, A. H.; Lee, K.; Cheng, M.; Wieteska, A.; Bartholomew, A. K.; Chen, Y. S.; Xu, X.; Pasupathy, A. N.; Zhu, X.; Dean, C. R.; Roy, X. Layered Antiferromagnetism Induces Large Negative Magnetoresistance in the van Der Waals Semiconductor CrSBr. *Adv. Mater.* **2020**, *32*, 2003240.

(28) Göser, O.; Paul, W.; Kahle, H. G. Magnetic Properties of CrSBr. *J. Magn. Magn. Mater.* **1990**, *92* , 129–136.

(29) Lee, K.; Dismukes, A. H.; Telford, E. J.; Wiscons, R. A.; Wang, J.; Xu, X.; Nuckolls, C.; Dean, C. R.; Roy, X.; Zhu, X. Magnetic Order and Symmetry in the 2D Semiconductor CrSBr. *Nano Lett.* **2021**, *21*, 3511–3517.

(30) Ghiasi, T. S.; Kaverzin, A. A.; Dismukes, A. H.; de Wal, D. K.; Roy, X.; van Wees, B. J. Electrical and Thermal Generation of Spin Currents by Magnetic Bilayer Graphene. *Nat. Nanotechnol.* **2021**, *16* , 788–794.

(31) Mangelsen, S.; Bensch, W. $HfTe_2$: Enhancing Magnetoresistance Properties by Improvement of the Crystal Growth Method. *Inorg. Chem.* **2020**, *59*, 1117–1124.

(32) Edelberg, D.; Rhodes, D.; Kerelsky, A.; Kim, B.; Wang, J.; Zangiabadi, A.; Kim, C.; Abhinandan, A.; Ardelean, J.; Scully, M.; others. Approaching the Intrinsic Limit in Transition Metal Diselenides via Point Defect Control. *Nano Lett.* **2019**, *19*, 4371–4379.

(33) Sung, S. H.; Schnitzer, N.; Brown, L.; Park, J.; Hovden, R. Stacking, Strain, and Twist in 2D Materials Quantified by 3D Electron Diffraction. *Phys. Rev. Mater.* **2019**, *3*, 064003.

(34) May, A. F.; Liu, Y.; Calder, S.; Parker, D. S.; Pandey, T.; Cakmak, E.; Cao, H.; Yan, J.; McGuire, M. A. Magnetic Order and Interactions in Ferrimagnetic $Mn_3Si_2Te_6$. *Phys. Rev. B* **2017**, *95,* 174440.

(35) Angelkort, J.; Wölfel, A.; Schönleber, A.; Van Smaalen, S.; Kremer, R. K. Observation of Strong Magnetoelastic Coupling in a First-Order Phase Transition of CrOCl. *Phys. Rev. B* **2009**, *80*, 144416

(36) Zhang, T.; Wang, Y.; Li, H.; Zhong, F.; Shi, J.; Wu, M.; Sun, Z.; Shen, W.; Wei, B.; Hu, W.; Liu, X.; Huang, L.; Hu, C.; Wang, Z.; Jiang, C.; Yang, S.; Zhang, Q. M.; Qu, Z. Magnetism and Optical Anisotropy in van Der Waals Antiferromagnetic Insulator CrOCl. *ACS Nano* **2019**, *13*, 11353–11362.

(37) Telford, E. J.; Dismukes, A. H.; Dudley, R. L.; Wiscons, R. A.; Lee, K.; Yu, J.; Shabani, S.; Scheie, A.; Watanabe, K.; Taniguchi, T.; Xiao, D.; Pasupathy, A. N.; Nuckolls, C.; Zhu, X.; Dean, C. R.; Roy, X. Hidden Low-Temperature Magnetic Order Revealed through Magnetotransport in Monolayer CrSBr. **2021**, arXiv:2106.08471.

(38) Ramirez, A. P.; Hayashi, A.; Cava, R. J.; Siddharthan, R.; Shastry, B. S. Zero-Point Entropy in "Spin Ice." *Nature* **1999**, *399*, 333–335.

(39) Zheng, J.; Ran, K.; Li, T.; Wang, J.; Wang, P.; Liu, B.; Liu, Z. X.; Normand, B.; Wen, J.; Yu, W. Gapless Spin Excitations in the Field-Induced Quantum Spin Liquid Phase of α-$RuCl_3$. *Phys. Rev. Lett.* **2017**, *119*, 227208.

(40) Liu, X.; Taddei, K. M.; Li, S.; Liu, W.; Dhale, N.; Kadado, R.; Berman, D.; Cruz, C. Dela; Lv, B. Canted Antiferromagnetism in the Quasi-One-Dimensional Iron Chalcogenide $BaFe_2Se_4$. *Phys. Rev. B* **2020**, *102,* 180403.